\documentclass[twocolumn,tighten]{aastex63}
\usepackage{amssymb, amsmath,framed}

\usepackage{bm}
\expandafter\ifx\csname package@font\endcsname\relax\else
 \expandafter\expandafter
 \expandafter\usepackage
 \expandafter\expandafter
 \expandafter{\csname package@font\endcsname}
\fi
\def\bq{\begin{equation}}
\def\eq{\end{equation}}
\def\bqy{\begin{eqnarray}}
\def\eqy{\end{eqnarray}}

\def\p{\partial}

\def\p{\partial}

\def\R{\mathbb{R}}

 \def\q#1#2{q^{\hspace{1pt}#1}_{\,,\hspace{1pt}#2}}
 \def\a#1#2{a^{\hspace{1pt}#1}_{\,,\hspace{1pt}#2}}
 \def\d#1#2{\delta^{\hspace{1pt}#1}_{\,,\hspace{1pt}#2}}

\begin{document}
\title{\large{A Bayesian Analysis of Technological Intelligence in Land and Oceans}}

\correspondingauthor{Manasvi Lingam}
\email{mlingam@fit.edu}

\author[0000-0002-2685-9417]{Manasvi Lingam}
\affiliation{Department of Aerospace, Physics and Space Sciences, Florida Institute of Technology, Melbourne, FL 32901, USA}
\affiliation{Department of Physics and Institute for Fusion Studies, The University of Texas at Austin, Austin, TX 78712, USA}

\author[0000-0002-3929-6932]{Amedeo Balbi}
\affiliation{Dipartimento di Fisica, Universit\`a di Roma ``Tor Vergata", 00133 Roma, Italy}

\author[0000-0003-2415-4840]{Swadesh M. Mahajan}
\affiliation{Department of Physics and Institute for Fusion Studies, The University of Texas at Austin, Austin, TX 78712, USA}
\affiliation{Department of Physics, School of Natural Sciences, Shiv Nadar University, Greater Noida, Uttar Pradesh, 201314, India}

\begin{abstract}
Current research indicates that (sub)surface ocean worlds essentially devoid of subaerial landmasses (e.g., continents) are common in the Milky Way, and that these worlds could host habitable conditions, thence raising the possibility that life and technological intelligence (TI) may arise in such aquatic settings. It is known, however, that TI on Earth (i.e., humans) arose on land. Motivated by these considerations, we present a Bayesian framework to assess the prospects for the emergence of TIs in land- and ocean-based habitats (LBHs and OBHs). If all factors are equally conducive for TIs to arise in LBHs and OBHs, we demonstrate that the evolution of TIs in LBHs (which includes humans) might have very low odds of roughly $1$-in-$10^3$ to $1$-in-$10^4$, thus outwardly contradicting the Copernican Principle. Hence, we elucidate three avenues whereby the Copernican Principle can be preserved: (i) the emergence rate of TIs is much lower in OBHs, (ii) the habitability interval for TIs is much shorter in OBHs, and (iii) only a small fraction of worlds with OBHs comprise appropriate conditions for effectuating TIs. We also briefly discuss methods for empirically falsifying our predictions, and comment on the feasibility of supporting TIs in aerial environments.\\
\end{abstract}

\section{Introduction}\label{SecIntro}
It is a well-known fact that liquid water (or ``water'' for short) is a critical prerequisite for life-as-we-know-it because it exhibits many desirable properties as a solvent \citep{PP12,PB17,SMI18}. It is not surprising, therefore, that the ``follow the water'' strategy is widely pursued in astrobiology \citep{HNG,MGK,WB18}; this approach is manifested, for instance, in the concept of the habitable zone (HZ) \citep{SHD,KWR93,KRK13,KRS14}.\footnote{The HZ, which embodies the region around the star where water could exist on a rocky planet's surface, has a long history dating back to at least the 19th century \citep{LM21}.} 

If we contemplate the Solar system, there are many worlds that host(ed) extensive bodies of liquid water. The majority of them can be termed ``ocean worlds'' (or ``water worlds'') due to the fact that they lack subaerial landmasses (notably continents). Earth itself appears to have been mostly devoid of subaerial landmasses for a fraction of its history as per several analyses \citep{HCD17}. In this context, some theoretical models and empirical geochemical constraints suggest that continents may not have emerged for most of the Archean \citep{FCR08,BZP18,JW20}, making the Earth effectively (although not completely) an ocean world during this interval of nearly $2$ Gyr since its formation.

In our Solar system, the vast majority of objects with liquid water may actually consist of \emph{subsurface oceans} underneath icy crusts, as opposed to containing oceans on the surface. The quintessential examples in this category are Enceladus and Europa, and several other worlds (e.g., Titan) are also confirmed to host subsurface oceans \citep{NP16,JL17,HHB19}. Hence, substantial attention has been devoted to gauging the habitability (via modeling and experiments) of the subsurface ocean worlds discovered in our Solar system \citep{HHH,TOV,CPG21,MBH21,GDG22}. The physicochemical prerequisites for habitability of planetary bodies, which is an inherently multi-faceted paradigm, are reviewed in \citet{LBC09,JK10,CBB16,CSS22,SBJ16,MaLi,SRK21}.

Looking beyond the Solar system, the discovery of debris disks (comprising analogs of the Kuiper belt) and exocomets \citep{MKW,HDM18,RVJ,SBK,REM20}, in a sizeable fraction of planetary systems \citep{MEK,SKW}, supports the surmise that small icy bodies -- which could host subsurface oceans of liquid water in principle -- are prevalent in the Milky Way. Changing tack, a combination of exoplanet observations and modeling has demonstrated that many of them are likely to be ocean worlds -- building on early proposals by \citet{MJK} and \citet{LSS04} -- with only oceans and no landmasses on the surface \citep{ZJS,ZJH,VGH,MDA,LP22,NLR22}. 

Some of the planets of the famous TRAPPIST-1 system detected in 2016-17 \citep{GTD} might belong to this group, and host water fractions of $\lesssim 10\%$ \citep{GDG,ADG21,ADM}; on the other hand, Earth's oceans merely add up to $\sim 0.02\%$ of its mass. The surficial habitability of ocean worlds has been explored in numerous publications \citep[e.g.,][]{CG15,KAG,NHR,KF18,MaL21,MPC21,SRI}. While the existence of liquid water is a given for these worlds (by their formulation), the prospects for maintaining stable climate, nutrient supply, and water-rock interactions, \emph{inter alia}, on geological timescales are less clear.

Yet, it is an incontrovertible datum that humans evolved on land-based habitats, and not in aquatic environments even though the latter are anticipated to be commonplace in the Milky Way.\footnote{In a similar vein, other seemingly unusual circumstances linked to humans include the presence of a large moon \citep{WB00,CRB01}; the location of Jupiter in the Solar system \citep{WB00,HJ08}; orbit around a G-type star \citep{HKW18,DK21}; and existence in the current cosmic epoch \citep{LBS16,DK21}.} Our emphasis on humans is not just mere anthropocentrism, but is instead a reflection of the canonical received notion that no other known species on Earth has (a) so radically transformed the biosphere in such a short timescale, even leading to the coinage of a potential new epoch, the Anthropocene \citep{ECE11,LM15,ZWW,WZ16}; and (b) generated signatures of its technology (technosignatures) that are detectable across significant (e.g., interstellar) distances \citep{JT01,SHW21}.

The preceding paragraphs raise the question: What is the probability of the emergence of technological intelligence (e.g., humans) in land- and ocean-based habitats? Earth is composed of both types of environments, but ocean worlds only possess the latter. This topic has an unexpectedly long history. Alfred Russel Wallace, who is renowned as the co-discoverer of the theory of natural selection (along with Charles Darwin), contended over a century ago that worlds with both continents and oceans are apposite for the evolution of complex animal-like lifeforms \citep[Chapter 12]{ARW03}. In recent times, by performing mathematical analyses, \citet{FS17} and \citet{MLin} posited that the concomitant existence of oceans and continents on Earth was pivotal for the genesis of technological intelligence.

In contrast, a few publications have postulated the opposite stance either implicitly or explicitly, to wit, that it is feasible to have technological intelligences emerge on ocean worlds, which consist of surface or subsurface oceans. As briefly reviewed hereafter in Section \ref{SSecRateIssue}, several marine animals evince multiple attributes of complex (higher-order) cognition. \citet{DLC18} conjectured, in connection with the Fermi paradox \citep{MMC18,DHF19}, that the majority of such intelligent organisms may be sealed in subsurface ocean worlds. It must be recognized, however, that in this case, the evolution of humans in a land-based environment would be rendered anomalous to a certain degree.

Motivated by the previous exposition, we carry out a quantitative (viz., Bayesian) analysis of the emergence of technological intelligence in land- and ocean-based habitats, taking our cue from Bayesian approaches in astrobiology that have sought to address a diverse array of unknowns \citep[e.g.,][]{DW11,DW17,ST12,BCL16,FS17,CKK18,DPW19,RDL19,BG20,DK20,DK21,SSD21,CHL22}. The structure of the paper is constructed based on the following line of reasoning.

We begin with clarifying some central terms employed throughout the paper and discussing the prevalence of worlds with LBHs and OBHs in Section \ref{SecPrelim}, thereby establishing the foundation(s) for the Bayesian analysis in Section \ref{SecMathFram}, which is mathematically (albeit not physically) equivalent to the formalism in \citet{DK21}; in this section, we also demonstrate how technological intelligences in land-based habitats may be anomalous. In Section \ref{SecCopAlt}, we qualitatively and quantitatively describe avenues whereby technological intelligences in land-based habitats could be rendered non-anomalous. Lastly, our salient findings are summarized in Section \ref{SecConc}.

\section{Basic characteristics of the model}\label{SecPrelim}
We will introduce a number of definitions and heuristics that are vital for our subsequent mathematical analysis. The reader may, instead, proceed directly to Section \ref{SecMathFram} for the statistical treatment.

\subsection{Model definitions}\label{SSecModDef}
To begin with, the classification that we shall tackle is the demarcation of land- and ocean-based habitats, because it is central to this work.\\

\noindent {\bf Land-based habitats (LBHs):} In land-based habitats, we include all terrestrial environments that can exist on worlds with landmasses, but \emph{not} on ocean worlds. In other words, this category encompasses not just subaerial settings like continents and volcanic islands (and the water bodies ensconced amid them), but also subterranean environments (typically) within the continental crust; the latter on Earth host the thriving deep biosphere \citep{EBC,MLD18}. A clarification worth underscoring is that organisms in land-based habitats would still require access to water, which is the solvent for life-as-we-know-it.\footnote{Alternative biochemistries may be viable \citep{Fir63,Ba04,BRC04,SMI18}, but remain empirically unsubstantiated to this date. Hence, we err on the side of caution and restrict ourselves to lifeforms predicated on the biochemistry of Earth (viz., carbon and water).} Moreover, in the limit of the land fraction approaching unity, the world is predicted to be covered by arid deserts with minimal biological productivity \citep{ARW03,MLin,HS22}. \\

\noindent {\bf Ocean-based habitats (OBHs):} By ocean-based habitats, we cover all environments that can theoretically exist on ocean worlds, as well as those harboring a mixture of oceans and landmasses (i.e., akin to Earth). Therefore, as indicated by the terminology, OBHs necessitate the presence of oceans in some form. This category includes putative environments on the seafloor (e.g., submarine hydrothermal vents) and underneath it (i.e., in the oceanic crust),\footnote{If the oceans (or overlying icy crusts) are deep and/or the parent worlds are large, the formation of high-pressure ices could prevent the actualization of some of these environments \citep{NHR,JK20,NM21}.} both of which feature diverse ecosystems on Earth \citep{WCW98,OSK11,WDO} in addition to the oceans themselves. The oceans can either occur on the surface or underneath an icy crust or ice-rock mixture \citep{VSK22}; the second case of this trio would resemble certain icy worlds (e.g., Enceladus) in our Solar system.\\

At first glimpse, these two eclectic categories appear to span the range of possible habitable worlds (for life-as-we-know-it) along with the myriad environments inherent to these worlds. It should be appreciated that OBHs and LBHs, although by their formulation cannot overlap, may nonetheless coexist in the same world, as is the case for Earth. A handful of other noteworthy subtleties warrant highlighting and elucidating. \\

\noindent 1. It is natural to wonder in which category amphibious organisms, which are quite widespread on Earth, should be assigned; as suggested by their nomenclature, the life cycles of these lifeforms may involve both LBHs and OBHs. By virtue of the manner in which the two classes were delineated, LBHs are automatically excluded from ocean worlds, which are devoid of landmasses altogether. If certain essential components of the life cycle (e.g., reproduction) of a particular species entail LBHs, it may be grouped in that category since LBHs still permit localized water bodies in which organisms might complete parts of their life cycle. In contrast, by simple tautology, OBHs are sans land environments. \\
By the same token, habitats that lie at the interface of landmasses and oceans are readily conceivable. We can seek to classify such habitats as either LBHs or OBHs on the basis of whether the underlying crust is continental or oceanic. If this distinction is not clear-cut, then we may categorize the environments based on which component (land or water) is more prominent in terms of area, volume, or some other salient characteristic. \\ 

\noindent 2. Aerial habitats are conspicuous by their absence hitherto. It is well-established that microbes survive in Earth's upper atmosphere \citep{DJS13,DD18}, and that the likes of Venus \citep{MS67,LMB21}, Jupiter and other gas giants \citep{Shap67,SS76}, and brown dwarfs \citep{Shap67,YPB17,ML19} might be capable of harboring aerial biospheres in principle. To the best of our knowledge, however, we are not cognizant of any organisms that complete their entire life cycle exclusively in Earth's atmosphere. More importantly, the subject of this study is technological intelligence (TI), as described below. We outline some feasible reasons in the Appendix as to why TI seems unlikely to transpire in aerial habitats.\footnote{It must be recognized that these reasons implicitly adopt Earth-based paradigms, which are not necessarily valid. However, in the absence of credible alternative pathways, it is a common strategy in astrobiology to employ Earth as the benchmark \citep[e.g.,][]{SB07,SJD15,CEC19,RP20}.} \\

\noindent 3. Although we have allowed for the possibility of subterranean and subseafloor habitats, respectively, within the continental and oceanic crust, we will discount such settings hereafter. The chief rationale is that the pores, cracks, and spaces in the rocks are conducive to the existence of microbes, but are anticipated to be inadequate for macroscopic organisms. As remarked in the preceding paragraph, we are interested in TI, and it is worth emphasizing that a general correlation between high cognition,\footnote{We caution, however, that constructing unequivocal metrics for bracketing organisms by their cognitive abilities, or a sliding scale for consciousness \citep{BSC20}, is riddled with problems.} brain size, and body size is documented on Earth \citep{Jer73,Arm83,Herc16}. In consequence, if the aforementioned environments can merely host microscopic (or mesoscopic) lifeforms, it is plausible that TI (needing high cognitive skills) would be untenable. Furthermore, these habitats are often severely energy-limited and/or nutrient-limited \citep{HJ13,LRL15,MLD18,BAA20}, thereby posing crucial issues for supporting TIs aside from the above constraint of available space. \\
In the same vein, while ice might be an appropriate medium for engendering the origin of life \citep{TSB,AWH} and/or hosting small extremophiles \citep{PBF07,MM10}, it does not appear suitable \emph{a priori} for hosting TIs, which are presumably macroscopic, because of the limited space, energy, and nutrients. Hence, ice-based habitats are excluded from our analysis of TIs. \\

Next, we unpack a vital aspect of the central theme of this paper, to wit, exploring the prospects of the emergence of lifeforms that belong to the same reference class as humans. Under what conditions, however, can organisms be placed in the same reference class as humans? In grappling with this question, we encounter a cognate fundamental question: What are the core differences between humans and nonhuman animals? Are they ``\emph{one of degree and not of kind}'', as succinctly posited by \citet[pg. 101]{CD71}? This subject has, unsurprisingly, attracted intense debate since at least the 19th century. Numerous publications contend that humans and nonhuman animals may be separated by a discontinuity or a profound gap in some respects \citep{PHP08,Cor11,Sud13,MT14,BC16,CH18,AK20}, while others favor the opposite stance vis-\`a-vis select traits or even on the whole \citep{Gri01,RD05,BP09,WR15,DW16,DW19,KA20}. When viewed in totality, the overall trend might be gradually shifting toward the latter camp, which builds upon the philosophy espoused by Darwin \citep{RGD21}.

In lieu of an in-depth discussion of this intricate and wide-ranging topic, it suffices to state that we will consider organisms occupying a socio-cognitive niche broadly analogous to that attributed to humans as belonging to the same reference class. This niche is endowed with components such as cultural transmission, language, and theory of mind \citep{SP10,BRH11,WE12,KL17}, \emph{inter alia}, and is predicated on a high level of technology in conjunction with the manifold facets of intelligence; or equivalently technological intelligence (TI). Therefore, we will designate the biological lifeforms drawn from this reference class as TIs, or sometimes as extraterrestrial technological intelligences (ETIs). While technology on the one hand and intelligence/cognition on the other are manifestly not independent -- in fact, they are deeply intertwined insofar as humans are concerned \citep{FE76,SLW59,SC12,OR20} -- the above focus on TI renders the connections with technosignatures more apparent.

Before moving on, we underscore that concepts such as ``technological intelligence'' and ``technology'' are subtle and exhibit a certain degree of ambiguity. Hence, it is conceivable that TIs in OBHs are endowed with characteristics that might place them in the same reference class as humans in some, but \emph{not} in all, important respects. As per the Ad Hoc Committee on SETI Nomenclature, (technological) ``intelligence'' may be understood as \citep{WSA18}:
\begin{quote}
In the acronyms SETI and ETI, the quality of being able to deliberately engineer technology which might be detectable using astronomical observation techniques.
\end{quote}
And ``technology'' was defined by the aforementioned committee to be \citep{WSA18}:
\begin{quote}
    The physical manifestations of deliberate engineering. That which produces a technosignature.
\end{quote}
Due to the complexity of these concepts, a detailed treatment lies beyond the scope of this paper.

\subsection{Model set-up}\label{SSecSetUp}
Although the quantitative analysis is pursued primarily in Section \ref{SecMathFram}, we will perform a couple of simple estimates herein that are employed later. 

In the ensuing calculations, the labels `$L$' and `$O$' signify LBHs and OBHs, respectively. Our goal is to gauge the potential number of worlds in the Milky Way with LBHs and OBHs that possess suitable conditions for the genesis and sustenance of TIs over geological timescales ($\gtrsim 1$ Gyr); these quantities are denoted by $N_L$ and $N_O$, respectively. In order to carry out these rough calculations, we will resort to a heuristic approach loosely reminiscent of the Drake equation \citep{Drake65,ISCS66}. Similar approaches have been adopted for addressing the origin of life \citep{SC16}, biosignatures \citep{SS18}, and technosignatures \citep{LL19,WH22}.

Let us begin with estimating $N_L$, because $N_O$ will be constructed similarly. Along the lines of the preceding publications, we will express $N_L$ as
\begin{equation}\label{NLdef}
    N_L \sim N_\star \cdot n_L \cdot f_L,
\end{equation}
where $N_\star$ is the total number of main-sequence stars in the Milky Way, $n_L$ is the mean number of worlds per main-sequence star that can host LBHs in principle, and $f_L$ is the fraction of such worlds that actually evince conditions appropriate for TIs, i.e., habitable in this sense. The factor $f_L$ encompasses multiple desiderata from habitability (aside from liquid water), abiogenesis, and TI. Likewise, the equation for $N_O$ is as follows:
\begin{equation}\label{NOdef}
    N_O \sim N_\star \cdot n_O \cdot f_O,
\end{equation}
where $n_O$ and $f_O$ are the oceanic counterparts of $n_L$ and $f_L$. We define the ratio of these two quantities by $\xi$, as it plays a major role later, and it simplifies to
\begin{equation}\label{Defxi}
    \xi \equiv \frac{N_L}{N_O} \sim \left(\frac{n_L}{n_O}\right) \left(\frac{f_L}{f_O}\right).
\end{equation}
Although $n_L$ and $n_O$ are unknown, it is still possible to constrain them (to an extent) because these variables fall under the purview of (exo)planetary science. 

Turning our attention to $n_L$, the worlds in question must be rocky and situated in the HZ (introduced in Section \ref{SecIntro}). The latter restriction follows from the fact that even surficial LBHs (see point \#1 in Section \ref{SSecModDef}) must have liquid water to permit habitability. The number of rocky (planet-sized) worlds in the HZ per star ($\eta_\oplus$) is not yet precisely determined. We will work with $\eta_\oplus \sim 0.1$ (see \citealt{LK17}), which could be slightly on the conservative side \citep{DC15,ZH19,BKK21}. Next, we must specify what fraction of these worlds may have a mixture of landmasses and oceans on the surface. Worlds devoid of surface water are excluded due to habitability issues and worlds sans land (i.e., ocean worlds) are excluded by definition in the context of estimating $n_L$.

Both theoretical estimates \citep[pg. 438]{ML21} and state-of-the-art simulations of exoplanets around low-mass stars \citep{KI22} indicate that the above fraction is $\lesssim 10\%$ \citep[cf.][]{TI15}. By combining the two factors, we end up with $n_L \sim 0.1 \times \eta_\oplus \sim 0.01$. The definition of $n_L$ delineated below (\ref{NLdef}) emphasizes that this estimate is merely in principle, and not in actuality. Hence, our calculation does \emph{not} imply that $1\%$ of all stars necessarily host rocky worlds with LBHs, and only suggests that they may do so. The rest of the uncertainty is, in essence, folded into the variable $f_L$; likewise, the corresponding uncertainty for OBHs is incorporated in $f_O$.

Now, we shall tackle $n_O$, where we must divide the calculation into two segments to account for surface and subsurface oceans. In the case of surface oceans, the analysis is similar to the previous paragraphs, which leads us to $\sim 0.35 \times \eta_\oplus$, with the fraction of $0.35$ drawn from the numerical simulations by \citet{KI22}; also refer to the model by \citet{TI15}. When it comes to subsurface oceans, however, the value of $n_O$ is much enhanced through two avenues. First, icy worlds with subsurface oceans are ostensibly common in the outer regions of planetary systems beyond the snow line. Second, a substantial number of icy planetesimals are ejected and comprise a free-floating population.

Tentative constraints on the number density of free-floating icy worlds can be derived by extrapolating microlensing studies to a certain size threshold \citep{SBM12,DG18}; the recent discovery of interstellar planetesimals has furnished additional data \citep{MM22,JS22}. With regard to icy worlds that are gravitationally bound to stars, their number is sensitive to planetary system architecture, stellar spectral type, and so forth, thereby requiring a statistical (population) study. Based on the available data, \citet{Man19} proposed that $n_O \sim 100$ for subsurface ocean worlds, which is compatible with the analysis of \citet{SJM21}. Both these publications allow for the possibility of $n_O \sim 1000$. We reiterate that $n_O$ denotes the number of worlds per main-sequence star that could support OHBs in principle.

After synthesizing the above data, and plugging them into (\ref{Defxi}), we duly end up with
\begin{equation}\label{XiFin}
    \xi \sim 10^{-4} \left(\frac{f_L}{f_O}\right),
\end{equation}
owing to which we adopt the fiducial value of $\xi \sim 10^{-4}$ henceforth, unless stated otherwise. This choice is obtained after specifying $f_L \sim f_O$ in (\ref{XiFin}). We revisit this vital assumption in Section \ref{SSecHabRare}, since it merely constitutes the default position whose validity is not assured. If the chosen value of $\xi \sim 10^{-4}$ is approximately correct, the potential number of worlds with OBHs (conducive to TIs) may outnumber those with LBHs by orders of magnitude, making the emergence of TIs in the latter conceivably anomalous, as discussed subsequently.

\section{Mathematical framework and implications}\label{SecMathFram}
With the various pieces assembled, we are equipped to expound our Bayesian approach. This formalism can be generalized, in principle, to a generic class of problems wherein some event/datum is observed in apparently unusual conditions associated with some reference class. However, as our current thrust is on TIs in LBHs and OBHs, we will tailor our exposition accordingly.

To recap from Section \ref{SSecSetUp}, the labels `$L$' and `$O$' stand for LBHs and OBHs, respectively. The variable $\xi$ embodies the ratio of the potential number of worlds with LBHs and OBHs with settings appropriate for TIs, and was estimated in (\ref{XiFin}). We are chiefly interested in calculating $P(L|\mathrm{TI})$ and $P(O|\mathrm{TI})$. The former roughly represents the probability of hosting TI (in the same reference class as humans) on LBHs, while the latter is the analogous probability for OBHs. To put it more precisely, each of these quantities embodies the probability that a randomly picked habitat belongs to the $L$ or $O$ category, given the condition that it hosts a TI.

As per the well-known Bayes's theorem \citep{HJ73,ETJ03}, $P(L|\mathrm{TI})$ can be expressed as
\begin{equation}\label{PLTI}
    P(L|\mathrm{TI}) = \frac{P(\mathrm{TI}|L)\,P(L)}{P(\mathrm{TI})},
\end{equation}
where $P(\mathrm{TI}|L)$ is the probability of the emergence of TI, given the condition that it unfolds in LBHs (namely, on worlds containing LBHs); $P(L)$ denotes the probability of selecting a world with LBHs; and $P(\mathrm{TI})$ is the probability of technological intelligence that will be defined shortly. Likewise, the formula for $P(O|\mathrm{TI})$ is derivable by replacing $L$ with $O$ in (\ref{PLTI}), thereby yielding
\begin{equation}\label{POTI}
    P(O|\mathrm{TI}) = \frac{P(\mathrm{TI}|O)\,P(O)}{P(\mathrm{TI})}.
\end{equation}

Computing the probabilities $P(L)$ and $P(O)$ is straightforward because this problem maps directly to the classic problem of drawing black and red balls from an urn. In consequence, these probabilities are given by
\begin{equation}\label{PL}
    P(L) = \frac{N_L}{N_L + N_O} = \frac{\xi}{\xi + 1},
\end{equation}
\begin{equation}\label{PO}
    P(O) = \frac{N_O}{N_L + N_O} = \frac{1}{\xi + 1},
\end{equation}
from which we notice that $P(L) + P(O) = 1$. This identity is expected because there are only two categories of habitats for TI in our treatment; note that aerial habitats are excluded, as adumbrated in Section \ref{SSecModDef}. It is worth recalling that habitable worlds could consist of both LBHs and OBHs (e.g., Earth) or just OBHs (ocean worlds), as per our conceptualization in Section \ref{SSecModDef}.

Now, we consider the conditional probability $P(\mathrm{TI}|L)$. If the origination of TI is encapsulated by a Poisson process \citep[e.g.,][]{BC83,ST12,SC16}, we can introduce a uniform rate parameter $\lambda_L$ for the emergence of TI, and a ``habitability interval'' $t_L$, which is the time over which TI could arise; both these parameters may be visualized as ensemble averages. Since $P(\mathrm{TI}|L)$ is the probability that at least one successful instantiation of TI has transpired in the habitability interval, for Poisson statistics we have
\begin{equation}\label{PTIL}
    P(\mathrm{TI}|L) = 1 - \exp\left(-\lambda_L t_L\right).
\end{equation}
In a similar vein, we can write down the equation for the conditional probability $P(\mathrm{TI}|O)$ as follows:
\begin{equation}\label{PTIO}
    P(\mathrm{TI}|O) = 1 - \exp\left(-\lambda_O t_O\right),
\end{equation}
where $\lambda_O$ and $t_O$ are the associated rate parameter and habitability interval for TI in OHBs, respectively.

Lastly, for computing $P(\mathrm{TI})$, we will utilize the law of total probability \citep[pg. 17]{AG05}, which leads to
\begin{equation}\label{PTI}
    P(\mathrm{TI}) = P(\mathrm{TI}|L) P(L) + P(\mathrm{TI}|O) P(O).
\end{equation}
Finally, substituting (\ref{PL})--(\ref{PTI}) into (\ref{PLTI}) and simplifying the ensuing expression, the probabilities are reduced to
\begin{equation}\label{PLTIFin}
    P(L|\mathrm{TI}) = \frac{\left[1 - \exp\left(-\lambda_L t_L\right)\right] \xi}{\left[1 - \exp\left(-\lambda_L t_L\right)\right] \xi + 1 - \exp\left(-\lambda_O t_O\right)},
\end{equation}
\begin{equation}\label{POTIFin}
    P(O|\mathrm{TI}) = \frac{1 - \exp\left(-\lambda_O t_O\right)}{\left[1 - \exp\left(-\lambda_L t_L\right)\right] \xi + 1 - \exp\left(-\lambda_O t_O\right)}.
\end{equation}
The two equations are formally equivalent to those derived in \citet{DK21}, who investigated the prospects for TIs on planets around M-dwarfs versus FGK stars. This exact correspondence is readily explainable from a mathematical standpoint because our work and \citet{DK21} both entail two categories of habitable worlds (albeit of different types) and explore the feasibility of the emergence of TI on such worlds. Of this duo, owing to the simple fact that human beings dwell on LBHs, the former is of more relevance and interest to us. We will, therefore, focus on (\ref{PLTIFin}) for the remainder of this section. It is apparent from (\ref{PLTIFin}) that there are five unknowns, of which we have tackled $\xi$ in Section \ref{SSecSetUp}, thence leaving us with four parameters. Of this quartet, we will now delve into the habitability intervals below.

As elucidated in Section \ref{SSecSetUp}, LBHs are linked with rocky worlds in the HZ. An upper bound on $t_L$ is consequently set by the amount of time that an object spends in the HZ. As per numerical models, the continuous HZ lifetime ranges from $\sim 6.8$ Gyr for $1\,M_\odot$ stars to $\sim 42$ Gyr for $0.2\,M_\odot$ stars \citep[Table 5]{RCO13}. However, it must be recognized that the habitability interval could be lower in actuality due to the negative effects of various physical processes. For instance, multispecies magnetohydrodynamic simulations have demonstrated that M-dwarf exoplanets with surface pressures of $\lesssim 1$ bar are susceptible to complete atmospheric depletion on timescales of $\lesssim 1$ Gyr because of substantial nonthermal escape rates \citep{GGD17,DLMC,DJL18,DHL19,DJL20,ABC20}.

Aside from the putative drawback of atmospheric retention, M-dwarf exoplanets are plausibly beset by a bevy of challenges such as tidal locking, prolonged pre-main-sequence phase with high irradiation, insufficient photon fluxes for prebiotic chemistry and photosynthesis, ozone depletion by regular stellar proton events, and frequency of (super)flares, to name a few \citep{TBM07,SBJ16,MaLi,ABC20}.\footnote{This group of factors may collectively explain why we do not find ourselves orbiting an M-dwarf \citep{DW11,HKW18,LM18,Ling19,DK21}.} The primary rationale for singling out M-dwarf exoplanets stems from the datum that M-dwarfs are the most common ($\sim 75\%$) and long-lived stars in the Milky Way \citep{TBM07}. Therefore, on account of all the earlier reasons, we adopt a somewhat conservative value of $t_L \sim 10$ Gyr in lieu of the suitable mean continuous HZ lifetime, which would be a few times higher than our current choice for $t_L$.

Next, we turn our gaze toward $t_O$. As suggested in Section \ref{SSecSetUp}, the vast majority of worlds with OBHs might be those with subsurface oceans based on extrapolation of known data. In estimating their abundance, we imposed a size cutoff roughly equal to the Moon (or Europa), motivated by the prediction that such worlds may retain liquid oceans on geological timescales of $\sim 1$ Gyr \citep{SS03,HSS06,SJM21}. It is conceivable, however, that the habitability interval for TIs is lower than this theoretical bound, and we will address this hypothesis further in Section \ref{SSecHabWTran}. At present, we work with the premise of $t_O \sim 1$ Gyr to explore the ramifications.

We are finally left with only two parameters, namely, the rate parameters $\lambda_L$ and $\lambda_O$. We will study the impact of varying rates in Section \ref{SSecRateIssue}, but we will regularly employ $\lambda_O \approx \lambda_L \equiv \lambda$ and seek to understand how $P(L|\mathrm{TI})$ behaves as a function of $\lambda$. Before doing so, we note that TI on Earth -- either genus \emph{Homo} in general or \emph{Homo sapiens} in particular -- emerged on an LBH in Africa roughly $4.4$ Gyr after the Earth first transitioned to certain habitable conditions such as water oceans and moderate temperatures \citep{WVP01,VPK02,TMH20}. Instead, if we presume that truly continuous habitable conditions were established only after an initial period of heavy bombardment, we may select a timescale of $\sim 4$ Gyr \citep{NS01}.

Thus, if we appeal to the so-called Principle of Mediocrity, also dubbed the Copernican Principle \citep[e.g.,][]{DD01,CS14}, we may tentatively suppose that $\lambda_L \sim 1/(4\,\mathrm{Gyr}) \sim 0.25\,\mathrm{Gyr}^{-1}$, which displays excellent agreement with the Bayesian analysis of \citet{DK21} that resulted in a median value of $\lambda_L \sim 0.26\,\mathrm{Gyr}^{-1}$. Hence, if a fiducial estimate for $\lambda_L$ is needed, we will accordingly adopt the latter value, although we emphasize that this rate is not tightly constrained.

\begin{figure}
\includegraphics[width=8.0cm]{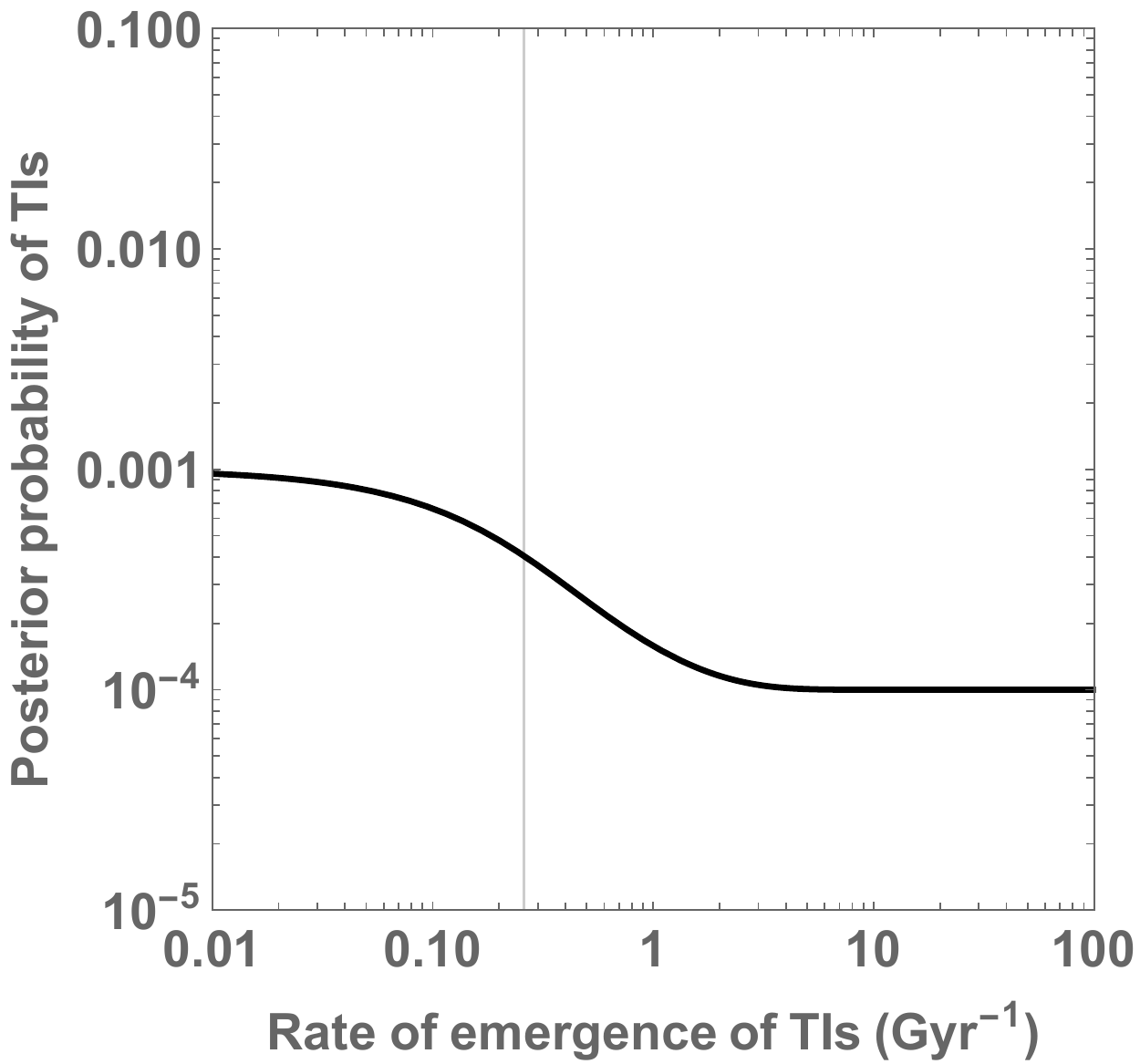} \\
\caption{The probability $P(L|\mathrm{TI})$ for TIs corresponding to LBHs as a function of the emergence rate of TIs in LBHs (units of Gyr$^{-1}$). The vertical line is an estimate of the median value of this rate, based on the data from Earth.}
\label{Posterior}
\end{figure}

On invoking the above simplifications, it is easy to verify that (\ref{PLTIFin}) is transformed into
\begin{equation}\label{PLTIFinv2}
    P(L|\mathrm{TI}) = \frac{\left[1 - \exp\left(-\lambda t_L\right)\right] \xi}{\left[1 - \exp\left(-\lambda t_L\right)\right] \xi + 1 - \exp\left(-\lambda t_O\right)},
\end{equation}
and before plotting this probability as a function of $\lambda$, it is instructive to evaluate two extreme cases. First, in the limit wherein $\lambda \rightarrow 0$ (i.e., the emergence of TI is exceptionally hard), we determine that (\ref{PLTIFinv2}) becomes
\begin{equation}\label{PLTILim1}
     P(L|\mathrm{TI}) \rightarrow \frac{t_L \xi}{t_L \xi + t_O} \sim 10^{-3},
\end{equation}
and taking the opposite limit of $\lambda \rightarrow \infty$ (viz., the genesis of TI is virtually guaranteed), we arrive at
\begin{equation}\label{PLTILim2}
     P(L|\mathrm{TI}) \rightarrow \frac{\xi}{\xi + 1} \sim 10^{-4}.
\end{equation}
On inspecting (\ref{PLTILim1}) and (\ref{PLTILim2}) and recalling that these two expressions act as bounds, it follows that $10^{-4} \lesssim  P(L|\mathrm{TI}) \lesssim 10^{-3}$, which is borne out by Figure \ref{Posterior}. Therefore, as the probability is always several orders of magnitude smaller than unity, we are led to the conclusion that the occurrence of humans (classified as TI) in LBHs is rendered highly atypical. To put it differently, we would crudely expect the majority of TIs to dwell in OBHs. 

If the preceding analysis is correct, then humanity's existence (in LBHs) would be apparently anomalous to a significant extent. This statement runs counter to the Principle of Mediocrity at first glimpse, which loosely posits that humankind is typical (viz., a random sample from an appropriate set) and that we are not special or privileged. We caution that there are manifold issues and subtleties associated with applying the Principle of Mediocrity \emph{tout court}, to wit, without taking the situation and context into proper consideration \citep{FC81,RM93,AK10,CB20}. Hence, it is not impossible that humans are indeed unusual in this particular respect (of originating in LBHs), while perhaps being typical in other ways. To put it more succinctly, humans might be representative of some reference classes and not of certain other reference classes. Thus, the basic concept of typicality is deeply intertwined with the choice of reference class. 

\section{Copernican alternatives}\label{SecCopAlt}
As the prior two paragraphs suggest, the ans{\"a}tzen we have chosen seem to collectively violate the Copernican Principle, which is often implicitly taken to hold true; this stance is potentially problematic, as indicated earlier. If we want to preserve the na\"ive version of the Copernican Principle, we must seek out tenable alternatives. In keeping with this theme, we shall refer to these hypotheses as ``Copernican alternatives''.

Because the final probabilities, (\ref{PLTIFin}) and (\ref{POTIFin}), are equivalent to those in \citet{DK21}, the Copernican alternatives that can be constructed are likewise equivalent, owing to which our discussion will largely mirror this reference. In general, we highlight that similar hypotheses may come into play as long as we are dealing with the emergence of life or TI in two generic categories that are, in essence, mutually exclusive and complementary. A necessary, although not strictly sufficient, criterion for Copernican alternatives is that $P(L|\mathrm{TI}) \gtrsim P(O|\mathrm{TI})$, thereby ensuring that the probability of TI linked with LBHs is higher than, or comparable to, that in OBHs. By employing (\ref{PLTIFin}) and (\ref{POTIFin}), this condition converts into
\begin{equation}\label{CopAlt}
    \left[1 - \exp\left(-\lambda_L t_L\right)\right] \xi \gtrsim 1 - \exp\left(-\lambda_O t_O\right),
\end{equation}
and we shall draw on this criterion hereafter. It is straightforward to extend this criterion to encompass a desired confidence level by introducing a prefactor (greater than unity) on the RHS.

At the outset, we emphasize that the trio of Copernican alternatives put forward are not, perforce, wholly independent of each other. Indeed, multiple processes and their effects may suppress the prospects for TIs in OBHs in tandem; these mechanisms could, in turn, overlap with more than one alternative. To offer a specific example, let us turn our attention to the wide-ranging domain of abiogenesis \citep{IF00,RBD14,Lu16,Suth17,PAB20}, and which sites (i.e., microenvironments) would be actually crucial or even imperative for facilitating the origin(s) of life \citep{SAB13,CDH19,SGS20,DCD22}.

If, for instance, LBHs such as hot springs \citep{DD20} or beaches and lagoons \citep{RM95,RL04} or arid intermountain valleys \citep{BKC12} are ineluctable for successful abiogenesis, this scenario would effectively disqualify OBHs, whereas including submarine alkaline hydrothermal vents -- widely, albeit by no means universally, perceived as promising environments for the origin of life \citep{MB08,RBB14,SHW16,MJR21} -- allow OBHs to instantiate the origin of life. The former case corresponds to $f_O \rightarrow 0$, and can have a bearing on both the characteristic emergence rate and habitability interval of TIs in OBHs.

By the same token, the trio of Copernican alternatives is not exhaustive. To single out a possible explanation, only TIs in LBHs may find themselves contemplating the question of the rate of occurrence of TIs in various habitats, and whether the Copernican Principle is violated in LBHs. In this example, however, it would be necessary to justify why most or all TIs in OBHs do not find themselves engaging with this question.

\subsection{Emergence rate of technological intelligence in OBHs is suppressed}\label{SSecRateIssue}
The first Copernican alternative we tackle entails subscribing to the notion that TI is much harder to engender in OBHs than in LBHs. This premise was intimated in \citet[Chapter 12]{ARW03}, and postulated in the Bayesian treatment by \citet{FS17}. If this conjecture is correct, the water-to-land transition of vertebrates constitutes one of the major evolutionary breakthroughs in Earth's history \citep{KB00}.

This proposition might seem untenable \emph{prima facie} because aquatic animals, especially cetaceans and cephalopods, evince manifold characteristics canonically ascribed to high cognition and intelligence \citep{MCF07,WR15,BSC20,SPBC}, such as mirror self-recognition (\citealt{RM01}, see also \citealt{KHT19}), self-control \citep{SBR21}, the ability to discriminate numbers \citep{AN21}, cultural transmission and creation of cultural niches \citep{WR15,FMS17}, complex communication \citep{JS13}, and tool use \citep{MP13}.

However, despite these attributes, it may be contended that none of the species dwelling in OBHs on Earth have evolved a level of TI exactly commensurate with that of humans, whereby the biosphere is shaped profoundly by their goal-directed actions. We will briefly describe some ostensibly plausible reasons that may hinder the emergence of TIs belonging to the same reference class as humans (refer to Section \ref{SSecModDef}) in OBHs, before embarking on a quantitative analysis of the criterion presented in (\ref{CopAlt}). Before doing so, we highlight that potentially $11$ out of $13$ high-performance innovations after the Ordovician transpired first (or only) in LBHs \citep{GV17}, and the majority of plant and animal biodiversity is documented in LBHs (consult \citealt{RMW22} and the references therein).

The density and viscosity of liquid water are $\sim 800$ and $\sim 50$ times higher than that of air (for a $1$ bar atmosphere), respectively. This fundamental datum suggests that the activity of organisms, broadly speaking, in OBHs may be limited by the medium of water relative to those in LBHs, which are typically expected to move in air \citep{MD93,AK20}. For starters, the drag force experienced by an organism is
\begin{equation}\label{FDrag}
    F_D \approx \frac{1}{2}C_D \mathcal{A}_\mathrm{org} \rho_f v^2,
\end{equation}
where $C_D$ is the drag coefficient, $\mathcal{A}_\mathrm{org}$ is the cross-sectional area of the organism, $\rho_f$ is the fluid density, and $v$ is the organismal velocity measured in the frame of the fluid. As long as the Reynolds number is of order unity and higher \citep[Section 7.8]{TEF95},\footnote{\url{https://www1.grc.nasa.gov/beginners-guide-to-aeronautics/drag-of-a-sphere/}} it follows that $F_D \propto \rho_f$ and by extension, the power needed to overcome the drag scales linearly with $\rho_f$. Therefore, \emph{ceteris paribus}, an organism moving through water is anticipated to consume $\sim 800$ times more energy than in air to offset the effect of drag.

Due to the higher power requirements, it is conceivable that putative budding TIs may be able to only traverse limited distances during their emergence, as doing otherwise would expend significant amounts of energy. These potential limitations on the home range (defined succinctly in \citealt{WHB43}) could have several consequences with respect to modulating intelligence. For instance, there is some evidence, albeit equivocal in nature, that home range correlates positively with certain aspects of cognitive capacity \citep{DNV,DWT03}; this correlation might be explainable by the necessity of possessing high cognition to handle the demands of navigating and utilizing a sizeable home range (\citealt{VE11}; see also \citealt{AGR17}).

Aside from the power constraint, we remark that swimming (intrinsically linked with OBHs) confers lower speeds in general compared to some modes of locomotion permitted in LBHs \citep{HJRB}. A simple scaling model developed by \citet{BM06}, which has subsequently been refined by later publications, determined that the ratio of optimal flying and running speeds on the one hand to swimming speeds on the other is $\left(\rho_\mathrm{org}/\rho_a\right)^{1/3}$, where $\rho_\mathrm{org}$ is the organism density and $\rho_a$ is the air density. After substituting the appropriate values (when $\rho_\mathrm{org}$ is close to that of water), the former duo (for Earth-like atmospheres) are about an order of magnitude higher than the swimming speeds.

One of the subtle, yet crucial, divergences between land and water concerns the scope for information gathering via sensory organs. This topic was reviewed in \citet{MWJ} and \citet{ABG16}, and its role was explicitly articulated in an astrobiological context by \citet{AK20} and \citet[Chapter 7.7.2]{ML21}. We shall first assess vision (with the proviso that the ambient radiation in a select wavelength range is sufficient for sensing purposes). The intensity $I$ at distance $d$ from the source is modeled as
\begin{equation}
    I = I_0 \exp\left(-\alpha d\right),
\end{equation}
where $I_0$ is the intensity at the source's location, and $\alpha$ is the attenuation coefficient. The attenuation coefficients in water and air are, respectively, $\sim 10^{-5}$ m$^{-1}$ and $\sim 10^{-1}$ m$^{-1}$ at $600$ nm; even at other similar wavelengths \citep[Figure 11.13]{MD93}, the discrepancy is $\gtrsim 2$-$3$ orders of magnitude. In turn, the optical visibility range of $\sim 80$ m in pure water \citep{MWJ} is over three orders of magnitude smaller than in clear air.\footnote{In contrast, sound waves are much less attenuated in water \citep[pg. 618]{ML21}, thus making them a promising avenue for information sensing and communication.}

The decreased visibility in water may have vital ramifications for the reification of high intelligence and TI, or lack thereof to be precise. The enhanced visual range on land could permit the evolution of more intricate predation strategies (relative to those in water) and concomitantly favor the emergence of complex responses from the prey, thereby possibly initiating a so-called evolutionary arms race. It has been theorized that the emergence of planning -- an important facet associated with higher-order cognition -- was facilitated by the water-to-land transition of vertebrates \citep{MSM17,MM20,MF22}. In the absence of LBHs, the evolution of this characteristic might not transpire at the same frequency. Before proceeding ahead, we reiterate that our exposition is not exhaustive. Subtle or prominent variations in the sensory modalities and \emph{Umwelten} of ``complex multicellularity'' \citep{AK11} dwelling in LBHs and OBHs, which are indirectly explored in \citet{JVU10} and \citet{EY22}, may translate to striking divergences in the probability of emergence of TIs in these environments.

If we focus on TI specifically, lifeforms in OBHs could be stymied by sparse access to raw materials and free energy sources to construct technology. With regard to the latter, to pick a potentially anthropocentric example, the development of fire control (by humans) was so pivotal that it has been postulated as part of one of the ``energy expansions'' in the evolutionary history of Earth (\citealt{Jud17}; see also \citealt{BBA,ACS18,SJP19}).\footnote{The timeline for the harnessing of fire by humans is subject to much uncertainty \citep{Gow16}, but it might have arisen as early as $\sim 1.5$ Ma \citep{HBF17,HCB19}.} From the narrower standpoint of human technology, fire patently offers numerous benefits (e.g., smelting of iron), but its relevance does not end there. Fire is believed to have contributed profoundly to hominin evolution in multiple ways: detoxifying food and boosting nutrient yield, constructing novel tools, keeping predators at bay, social bonding near the hearth, and expanding into new and otherwise inhospitable environments \citep{BBA,Wra09,Wra17,VS08,VS10,VS17,ACS18,SJP19}.

Subsurface ocean worlds seemingly comprise the most common repositories of OBHs, as outlined in Section \ref{SSecSetUp}. Hence, in many (and perhaps most) OBHs, it is plausible that fire could be absent if one or more of the fuel, oxidant, or heat source is unavailable, thereupon posing major hurdles to reifying TI. With that being said, if we specialize to Earth, the temperature of certain submarine hydrothermal vents can reach $> 400^\circ$C \citep{HPK07,KG08} and the melts of submarine volcanoes possess initial temperatures well above $1000^\circ$C \citep[pg. 390]{KBD}, thus constituting plausible energy sources that may be harnessed in lieu of fire. Analogs of fire (vis-\`a-vis providing substantial sources of free energy and temperature) could exist in OBHs if strong oxidants and reductants are colocated, although no such concrete alternatives have been identified so far. Lastly, with respect to raw materials for creating technologies, extracting them from the ocean subseafloor (under high pressures) and transporting them against the drag might entail unforeseen challenges.

\begin{figure}
\includegraphics[width=8.0cm]{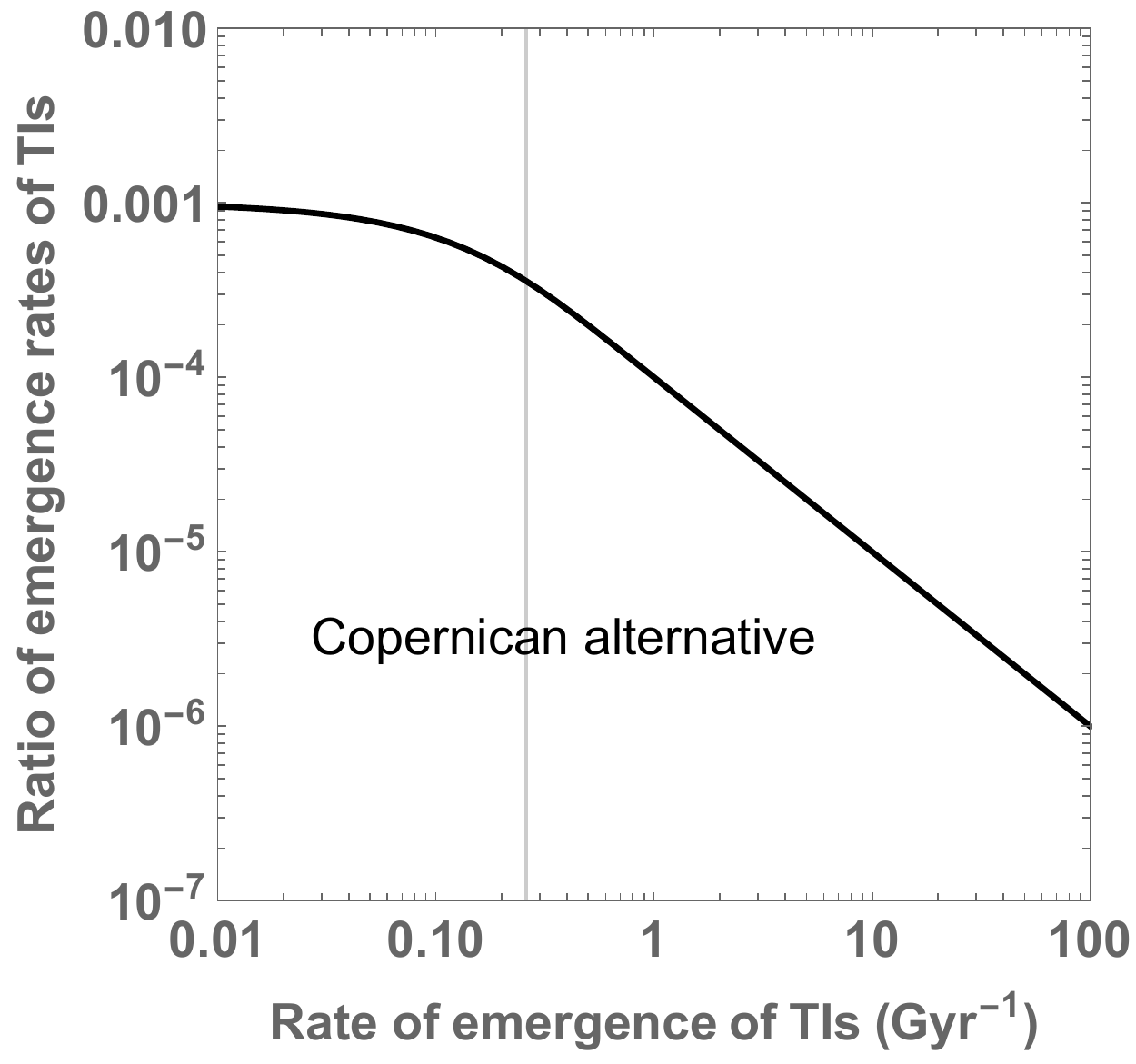} \\
\caption{Ratio of the emergence rates of TIs in OBHs and LBHs as a function of the rate of emergence of TIs in LBHs (units of Gyr$^{-1}$) to achieve a resolution roughly compatible with the Copernican principle. The region under the curve spans the parameter space for this Copernican alternative. The vertical line is an estimate of the median value of the emergence rate, based on the data from Earth.}
\label{RatioRate}
\end{figure}

Hitherto, we have delineated several mechanisms that may be responsible for suppressing the emergence of TI in OBHs. Now, we will compute the quantitative outcomes of (\ref{CopAlt}). Our goal here is to determine the constraint(s) on $\lambda_O$ while holding all parameters aside from $\lambda_L$ fixed at their default values. On solving for $\lambda_O$ as a function of $\lambda_L$, we end up with
\begin{equation}\label{LOCons}
    \lambda_O \lesssim -\frac{\ln\left(1 - \xi\left[1 - \exp\left(-\lambda_L t_L\right)\right]\right)}{t_O}.
\end{equation}
If we take the limit of $\lambda_L \rightarrow \infty$ (i.e., TI is common in LBHs) and $\xi \ll 1$, we obtain the analytical expression 
\begin{equation}\label{Lim1LO}
   \lambda_O \lesssim \frac{\xi}{t_O}. 
\end{equation}
On the other hand, taking the opposite limit of $\lambda_L \rightarrow 0$ (i.e., TI is rare in LBHs) and $\xi \ll 1$, we arrive at
\begin{equation}\label{Lim2LO}
   \lambda_O \lesssim \frac{\xi\,\lambda_L t_L}{t_O}
\end{equation}
Between (\ref{Lim1LO}) and (\ref{Lim2LO}), the smaller of the duo is the latter because we relied on the condition $\lambda_L t_L \ll 1$. In other words, we have derived an upper bound on the ratio of the rate parameters that is given by
\begin{equation}
    \frac{\lambda_O}{\lambda_L} \lesssim \xi\, \frac{t_L}{t_O} \lesssim 10^{-3},
\end{equation}
where the last equality has followed after substituting the fiducial values for the parameters on the RHS. Hence, if the rate of the emergence of TIs in OBHs is at least three orders of magnitude smaller than the corresponding rate in LBHs, this explanation could serve as a viable Copernican alternative. We have plotted the ratio $\lambda_O/\lambda_L$ as a function of $\lambda_L$, which is calculated from (\ref{LOCons}, in Figure \ref{RatioRate} and it is observed that the above trends manifest along expected lines.

\subsection{Habitability interval in OBHs is transient}\label{SSecHabWTran}
In the second Copernican alternative, we will specify $\lambda_O \approx \lambda_L$ in the same vein as Section \ref{SecMathFram} and counter to that of Section \ref{SSecRateIssue}. Instead, the hypothesis we investigate is that the habitability interval for TI in OBHs (which is encapsulated by $t_O$) is suppressed by orders of magnitude with respect to LBHs. Before tackling the mathematical aspects, we outline a couple of credible routes that may engender this scenario.

It is plausible and perhaps likely that TIs are predicated on metabolic pathways that are highly exergonic, and yield ample energy to carry out essential functions such as growth, maintenance, and reproduction. If this reasonable premise is correct, it automatically follows that the substrates involved in these pathways must be available in sufficient abundances, among other prerequisites for the existence of TIs. As a corollary, in the event that OBHs permit such metabolisms to operate only for transient periods of time on the whole, this trend would lower the prospects for TIs in OBHs and therefore act as a tenable Copernican alternative.

Let us consider a widely studied example in our Solar system: Enceladus. A combination of theoretical models, laboratory experiments, and observational data from the \emph{Cassini} mission appear to support the conclusion that the timescale for serpentinization on Enceladus is $\sim 100$ Myr \citep{ZTH,DCS22}. Of the noteworthy byproducts of serpentinization, which has been posited as ``\emph{life's mother engine}'' \citep{RNB13}, molecular hydrogen (H$_2$) stands out since it can be employed in ancient chemoautotrophic pathways such as hydrogenotrophic methanogenesis and acetogenesis \citep{CSC20}. Hence, if the availability of H$_2$ were to be temporally curtailed in general (more so than in Enceladus), then the probability of the emergence of life and possibly TIs in OBHs may be diminished.

Second, let us contemplate one of the most potent oxidizing agents in metabolic pathways on Earth: molecular oxygen (O$_2$). In worlds with subsurface oceans, which are anticipated to constitute the dominant repositories of OBHs (refer to Section \ref{SSecSetUp}), O$_2$ levels are modulated by the delivery of oxidants from the surface and by the radiolysis of water, among other channels \citep[e.g.,][]{CH01,CP01,HCC07,BGW,RMH,Man19,RGW}; aside from the sources, the magnitude(s) of the sinks must be taken into account in gauging the O$_2$ abundance. If the dissolved oxygen attains high-enough concentrations, it is plausible that OBHs in subsurface ocean worlds could support TIs, by virtue of the fact that aerobic metabolisms can yield approximately an order of magnitude more energy than anaerobic metabolisms for the same food intake \citep{CGZ05,TM07,KB08}.

At this stage, a brief digression is warranted. The preceding exposition may suggest at first glimpse that once the O$_2$ levels exceed a certain threshold, the advent of TIs would be facilitated. In actuality, the existence of such a threshold is hard to establish, if the Earth's record is anything to go by. The hypothesis that the evolution of animals (to which humans as TI belong) is directly linked to the rise in Earth's atmospheric O$_2$ has a long history \citep{JRN59,PC68,BR82}, but this posited causality has been called into question by recent empirical evidence \citep{CME20,LDP21,MBD22,SBD22}; in particular, some animals are documented to readily survive under low-oxygen (e.g., \emph{Demospongiae}) or completely anoxic (e.g., \emph{Henneguya salminicola}) conditions \citep{DDP10,MWJ14,YAN20}. Notwithstanding this caveat, it is still conceivable that the evolution of the forerunners of TIs (e.g., complex multicellular lifeforms that are macroscopic and motile) might necessitate high O$_2$ concentrations \citep{MKJ23}; the latter could also promote higher biodiversity and complex food webs \citep{LG98,SFR13,KS14}.

Circling back to the original theme, if worlds with OBHs accrue substantial O$_2$ levels only after a lengthy duration (labeled by $t_\mathrm{oxy}$), this path might translate to a truncated habitability window for the emergence of TIs. The timescale for oceanic oxygenation could be $\gtrsim 1$ Gyr \citep{RG10,ML21}, consequently narrowing the habitability window in principle. To illustrate how this sequence of events may occur, we must recognize that the depth of the subsurface ocean can decrease over time as the icy crust thickens with a decline in the internal heat budget, eventually freezing over; the timescale is denoted by $t_\mathrm{freeze}$. The habitability interval would thus be $\sim \left(t_\mathrm{freeze} - t_\mathrm{oxy}\right)$ in this context, which ought to become small if $t_\mathrm{oxy} \approx t_\mathrm{freeze}$ is valid or even zero if $t_\mathrm{oxy} \geq t_\mathrm{freeze}$. To sum up the previous paragraphs, temporal availability of reducing and oxidizing agents represents a potential method of decreasing $t_O$.

\begin{figure}
\includegraphics[width=8.0cm]{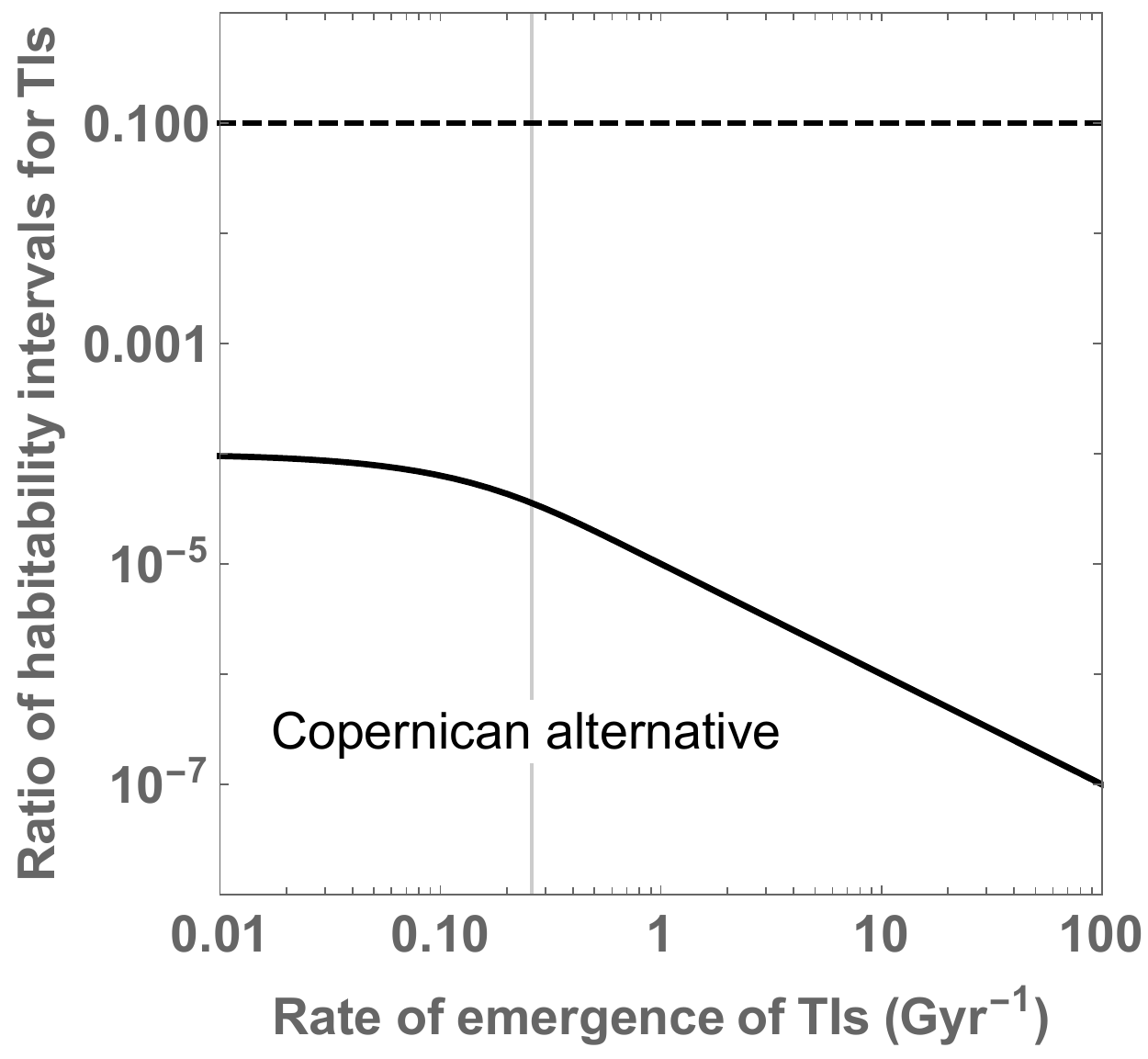} \\
\caption{Ratio of the habitability intervals for TIs in OBHs and LBHs as a function of the emergence rate of TIs in LBHs (units of Gyr$^{-1}$) to achieve a resolution roughly compatible with the Copernican principle. The region under the curve depicts the parameter space for this Copernican alternative. The vertical line is an estimate of the median value of the emergence rate, based on the data from Earth. The horizontal dashed line is the fiducial value of this ratio, which is substantially greater than the Copernican alternative.}
\label{RatioTime}
\end{figure}

We will now perform our mathematical analysis; before doing so, recall that we have set $\lambda_O \approx \lambda_L \equiv \lambda$. On substituting this expression in (\ref{CopAlt}), and solving for $t_O$ as a function of $\lambda$, we accordingly obtain
\begin{equation}\label{TOCons}
    t_O \lesssim -\frac{\ln\left(1 - \xi\left[1 - \exp\left(-\lambda t_L\right)\right]\right)}{\lambda}.
\end{equation}
Before undertaking the plot, it is instructive to calculate the twin limits of $\lambda \rightarrow \infty$ and $\lambda \rightarrow 0$ after using the relation $\xi \ll 1$. The former case leads us to
\begin{equation}\label{Lim1tO}
   t_O \lesssim \frac{\xi}{\lambda}, 
\end{equation}
and determining the second limit yields
\begin{equation}\label{Lim2tO}
   t_O \lesssim \frac{\xi \,\lambda t_L}{\lambda},
\end{equation}
and of these two equations, the latter is the relevant bound because of the built-in ordering of $\lambda t_L \rightarrow 0$. Thus, after simplifying (\ref{Lim2tO}), we arrive at
\begin{equation}
    \frac{t_O}{t_L} \lesssim \xi\ \lesssim 10^{-4},
\end{equation}
after we have utilized the default choice for $\xi$. The above equation tells us that $t_O$ must be lowered by at least four orders of magnitude relative to $t_L$. Since we have chosen $t_L \sim 10$ Gyr, we would require $t_O \lesssim 1$ Myr; in contrast, observe that the fiducial value for $t_O$ is much higher at $\sim 1$ Gyr. Hence, the factor by which $t_O$ must be suppressed is undoubtedly significant, and it is unclear as to whether the premise outlined in this subsection is entirely tenable for this reason. We have plotted $t_O/t_L$ as a function of $\lambda$ in Figure \ref{RatioTime}, from which we see that the aforementioned characteristics are all apparent. For example, it is evident that $t_O/t_L$ must be suppressed by orders of magnitude relative to its standard value.

Lastly, we comment on the fact that we held $t_L$ fixed and sought to examine the ensuing constraints on $t_O$. In reality, the chief quantity of interest to us is the ratio $t_O/t_L$. Therefore, boosting $t_L$ could also count as a feasible Copernican alternative, given that it would be analogous to decreasing $t_O$ instead. The elevation of $t_L$ may arise if LBHs predominantly occur on K- and M-dwarf exoplanets, and the habitability intervals of these worlds are markedly higher than their HZ lifetimes.\footnote{In the domain of technosignatures, we point out that such worlds have been contemplated \citep{Cri85,GLM17}, although they crucially presuppose the existence of TIs in the first place.}

\subsection{Fraction of habitable worlds with OBHs is low}\label{SSecHabRare}
For TIs to originate, it is virtually a tautology to say that the worlds in question must have some basic set(s) of conditions that allow for this possibility. This notion of ``habitability'' in connection with TIs is encapsulated in the factor $\xi$, specifically in the form of $f_L$ and $f_O$, as delineated in Section \ref{SSecSetUp}. Hence, if we relax the assumption made hitherto, namely that $f_L \sim f_O$, it is easy to enhance $\xi$ and thereby fulfill the criterion in (\ref{CopAlt}). Hence, the third Copernican alternative we evaluate postulates that the majority of worlds with OBHs are outright not suitable for TIs (and their emergence) in any fashion. 

As before, it behooves us to identify potential mechanisms that may suppress $f_O$ and boost $\xi$. Although numerous avenues could exist, we will single out a couple of candidates. The first pertains to the access to energy sources and the next two revolve around the abundances of vital nutrients, all of which can be a major hurdle for OBHs in subsurface ocean worlds; recall that this category of worlds is predicted to be particularly common in the Galaxy, as reviewed in Section \ref{SSecSetUp}.

We have touched on the first theme in both Sections \ref{SSecRateIssue} and \ref{SSecHabWTran}. In the former, we discussed how certain OBHs may lack energy sources that can be efficiently deployed by TIs, akin to the role(s) played by fire on Earth with regard to humans. In the latter, we commented on the centrality of molecular oxygen for supporting complex life and TIs. It should be recognized in this context that not all ocean worlds are likely to attain high levels of dissolved O$_2$ if we extrapolate from the Solar system \citep{WSH}, and in theory, only a small fraction of them might actually do so. For instance, it is plausible that a sizeable fraction of worlds with (sub)surface oceans lack O$_2$ concentrations adequate for organisms resembling macroscopic motile animals \citep{MLin,GHD,HS22}.

The second major justifiable impediment is nutrient availability. A bevy of publications have proposed that a subset of worlds with surface \citep{WP13,MLin,GHD,OJA} or subsurface \citep{MYZ,ML18} oceans may evince a scarcity of dissolved phosphorus (a bioessential element for life-as-we-know-it) in the form of phosphates, while other studies have elucidated avenues that could raise phosphorus concentrations to Earth-like levels or higher \citep{PHB,SRI,BTT,HGH,POF}. Looking beyond phosphorus, many other elements are critical for life-as-we-know-it \citep{DSW01,WDE04}, and even if one or a few of them are scarce, putative complex biospheres might be ruled out. Thus, when viewed cumulatively, the dual restrictions imposed by the need for appropriate energy sources and nutrients can jointly suppress $f_O$ to produce the desired outcome.

Third, if worlds with OBHs are widely endowed with substantial H$_2$O inventories, manifesting as thick ice layers (for subsurface ocean worlds) and/or deep oceans, the pressures at the ocean floor can become high enough (typically $\gtrsim 1$ GPa) to drive the formation of high-pressure ices \citep{PW99}. In this scenario, vital water-rock reactions might be suppressed, thereby stymieing access to nutrients, substrates, chemical energy, and miscellaneous sources of disequilibria \citep{NHR,JK20,BJ22}. However, this drawback may be mitigated by recent numerical models, which imply that the slow transport of salts, nutrients, and other substances into the ocean is feasible even when high-pressure ices exist \citep{CTS,JDP,KS18,HCL,OTB}.

\begin{figure}
\includegraphics[width=8.0cm]{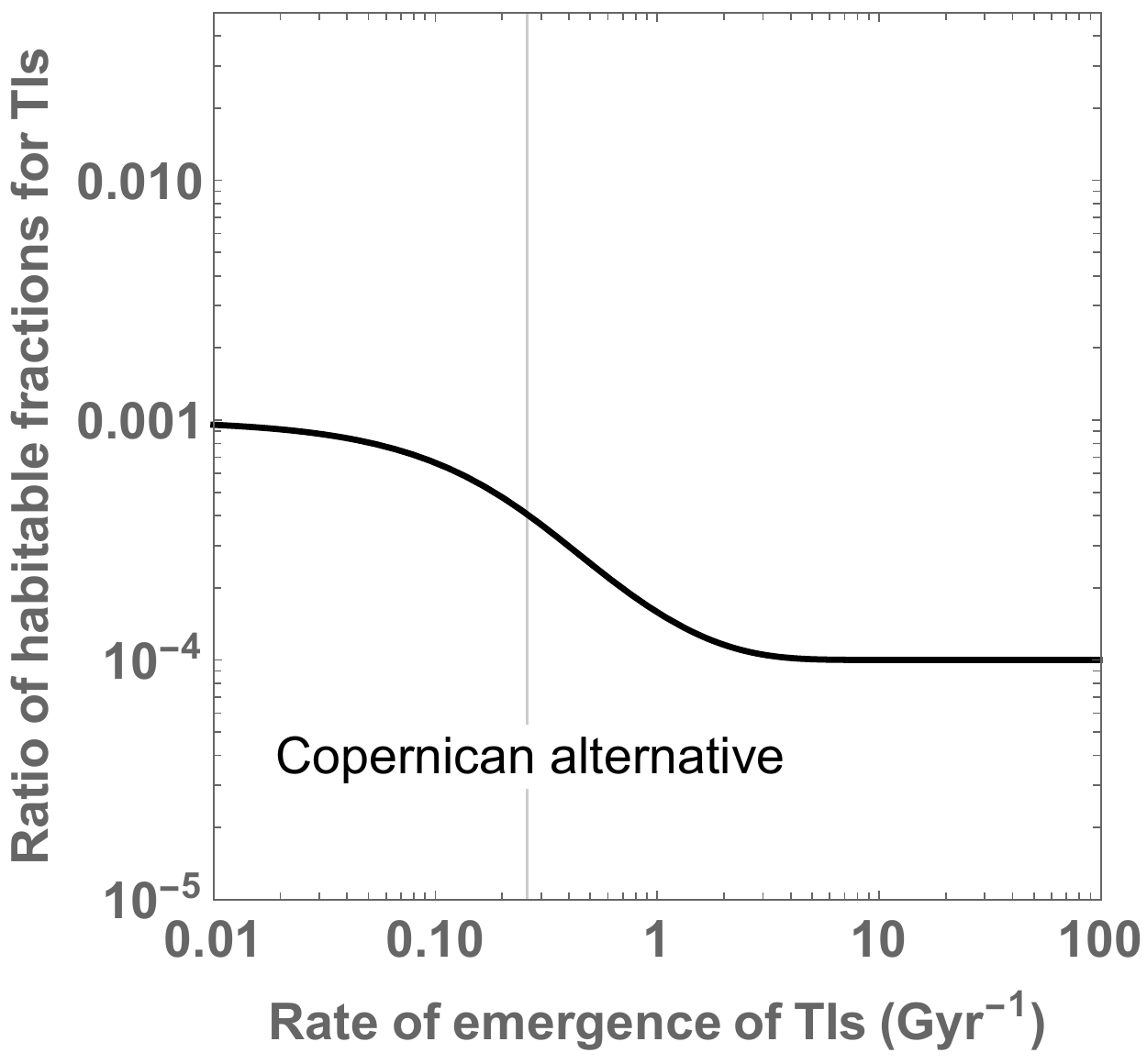} \\
\caption{Ratio of the fractions of habitable worlds (for TIs) with OBHs and LBHs as a function of the emergence rate of TIs in LBHs (units of Gyr$^{-1}$) to achieve a resolution roughly compatible with the Copernican principle. The region under the curve is the parameter space for this Copernican alternative. The vertical line is an estimate of the median value of the emergence rate, based on the data from Earth.}
\label{RatioHab}
\end{figure}

Returning to (\ref{CopAlt}), we can derive the necessary lower bound for $\xi$, which duly leads to
\begin{equation}\label{xiCond}
     \xi \gtrsim \frac{1 - \exp\left(-\lambda t_O\right)}{1 - \exp\left(-\lambda t_L\right)},
\end{equation}
and we have imposed the previous ordering $\lambda_O \approx \lambda_L \equiv \lambda$ because our goal is to determine $\xi$ as a function of $\lambda$. Let us first take the limit of $\lambda \rightarrow \infty$, which corresponds to the emergence of TIs being virtually guaranteed -- we end up with the simple expression $\xi \gtrsim 1$. On substituting (\ref{XiFin}) into $\xi \gtrsim 1$ and simplifying, we arrive at 
\begin{equation}\label{RatEas}
    \frac{f_O}{f_L} \lesssim 10^{-4}.
\end{equation}
Now, let us tackle the opposite regime wherein the emergence of TI is extremely hard ($\lambda \rightarrow 0$), which yields
\begin{equation}\label{xiLB}
     \xi \gtrsim \frac{t_O}{t_L} \gtrsim 0.1,
\end{equation}
where the second inequality follows from inputting the fiducial values for $t_O$ and $t_L$ motivated in Section \ref{SecMathFram}. After plugging (\ref{XiFin}) into (\ref{xiLB}), we arrive at 
\begin{equation}\label{RatHard}
    \frac{f_O}{f_L} \lesssim 10^{-3}.
\end{equation}
Therefore, upon inspecting (\ref{RatEas}) and (\ref{RatHard}), it is apparent that the fraction of habitable worlds (\emph{sensu} capable of engendering TIs) with OBHs must be suppressed by approximately $3$-$4$ orders of magnitude at the minimum compared to LBHs. Even though we have highlighted some bottlenecks to habitability with respect to TIs in OBHs, this factor is undoubtedly substantial, indicating that the vast majority of worlds (typically with subsurface oceans) comprising viable OBHs must be fully incompatible with the conditions required for TIs. 

Instead of plotting $\xi$, we have opted to depict the above ratio $f_O/f_L$ in Figure \ref{RatioHab}. This quantity is readily obtained by merging (\ref{XiFin}) and (\ref{xiCond}), and we obtain 
\begin{equation}
      \frac{f_O}{f_L} \lesssim 10^{-4} \left[\frac{1 - \exp\left(-\lambda t_L\right)}{1 - \exp\left(-\lambda t_O\right)}\right].
\end{equation}
Upon perusing Figure \ref{RatioHab}, the analytical criteria embodied by (\ref{RatEas}) and (\ref{RatHard}) are manifested as anticipated.

In closing, let us recall that the current objective was to boost $\xi$ compared to its fiducial value of $\sim 10^{-4}$. We have sought mechanisms that suppress $f_O/f_L$, and thus raise $f_L/f_O$ and $\xi$. On examining (\ref{Defxi}), however, we notice that $\xi$ can be increased if $n_L/n_O$ is enhanced. This result is realizable if the number of worlds with the basic \emph{potential} to host LBHs is higher than expected, or the opposite trend (i.e., lower) is applicable for OBHs. Exoplanet surveys and theoretical models have already shed light on $n_L/n_O$, and this quantity will be further resolved with additional data in the future.

\section{Conclusion}\label{SecConc}
Humans are an example of technological intelligence (TI), albeit of the specific kind that can profoundly influence the biosphere through their purposeful activities and produce detectable signatures of their technology. It is a well-established fact that TI on Earth arose on land, and not in the oceans, despite the prediction that ocean worlds should be prevalent in the Milky Way. In this paper, we performed a Bayesian analysis of the probability of TIs existing in LBHs and OBHs.

There are four broad outcomes that appear to be consistent with the datum that TI on Earth emerged in a particular LBH; of this quartet, the first seemingly violates the Copernican Principle, while the other trio ostensibly preserve the elementary form of this principle.
\begin{enumerate}
    \item The existence of TI in LBHs on Earth is a genuine ``fluke'' with odds ranging from $1$-in-$10^3$ to $1$-in-$10^4$. To put it another way, one would expect the overwhelming majority of TIs to inhabit OBHs (which does not seem compatible with available data from the Solar system).
    \item OBHs have a much lower (ensemble-averaged) rate of emergence of TIs relative to their counterpart for LBHs. To be precise, in case the rate associated with OBHs is at least three orders of magnitude smaller than that of LBHs, the fact that TI dwells in LBHs on Earth is not anomalous.
    \item OBHs are endowed with a much more transient interval of habitability for TIs compared to LBHs. If the (ensemble-averaged) habitability timescale for OBHs is more than four orders of magnitude lower than its analog for LBHs, the existence of TI on Earth in LBHs would not be anomalous.
    \item Only a minuscule fraction of worlds with OBHs (with respect to LBHs) feature desiderata conducive to the emergence of TIs. If the fraction of worlds containing OBHs with habitable conditions for TIs is smaller by $3$-$4$ orders of magnitude than the corresponding fraction for LBHs, the presence of TI in LBHs of Earth is not anomalous.
\end{enumerate}
As remarked above, a clear distinction between hypothesis \#1 and the remaining three possibilities arises automatically. In consequence, we are naturally propelled toward the question of how to differentiate between these scenarios, and thence falsify or validate them.

Let us contemplate the first conjecture once more. If OBHs are not significantly disfavored in some fashion compared to LBHs, then TIs should be far more common in the former. In this context, determining whether a world has surficial landmasses and/or oceans is feasible, in principle, via spectrophotometric observations \citep{CAM09,FKS10,FAD18,LMT18,KKA22}. Hence, if future technosignature surveys (reviewed in \citealt{JTW21,ML21,SHW21,HSS22}) discover that the majority of signals emanate from ocean worlds (sans LBHs by definition), this trend might assist in confirming hypothesis \#1. In contrast, if most technosignatures originate from worlds with LBHs, this result may serve to falsify hypothesis \#1 and thereby lend credence to the other outcomes. We caution, however, that the process of falsification and verification is not straightforward as the likes of false positives and negatives must be accurately addressed.

For the sake of argument, let us suppose that we have ruled out hypothesis \#1 as described in the prior paragraph. This route would still leave us with the conundrum of determining which of hypotheses \#2, \#3, and \#4 is/are correct. In view of the rather limited scope of surveys as well as the data garnered from them in the near-future, it seems very unlikely these hypotheses can be differentiated from one another. Thus, at least in the upcoming decades, any progress on this front could be restricted to performing careful extrapolations from Earth and/or carrying out theoretical modeling.

In summary, we have tackled the fundamental question of why we -- in the specific sense of constituting a TI -- find ourselves having evolved in LBHs and not in OBHs, despite the latter being potentially much more common than the former. A Bayesian approach suggests that our emergence in the former setting was indeed deeply unlikely \emph{prima facie}, unless certain mechanisms act to selectively suppress the prospects for TIs in OBHs relative to LBHs. Future surveys for technosignatures, backed by forthcoming missions seeking biosignatures, may shed welcome empirical light on this question, and enable us to gauge whether TIs in LBHs are actually uncommon (in comparison to OBHs).

\acknowledgments
M.L. is grateful to Andy Knoll for several illuminating and thought-provoking conversations over the years concerning related subjects, and the CATS (Characterizing Atmospheric Technosignatures) collaboration for stimulating group meetings. 

A.B. acknowledges support by grant numbers FQXi-MGA-1801 and FQXi-MGB-1924 from the Foundational Questions Institute and Fetzer Franklin Fund, a donor-advised fund of the Silicon Valley Community. M.L. acknowledges partial support from the NASA Exobiology program under grant 80NSSC22K1009.

\appendix

As stated in Section \ref{SSecModDef}, the objective of this Appendix is to sketch severe challenges that could confront the emergence of TIs in aerial biospheres, which may explain why atmospheric settings are unsuited for this purpose. We emphasize that this summary is not exhaustive since other drawbacks can be readily identified.

The first hurdle we wish to underscore is the abundance of bioessential elements in the atmosphere. Even if the availability of certain lighter bioessential elements and compounds (e.g., carbon and water) does not comprise a bottleneck (a premise that is, however, not assured), the situation could prove to be completely different when it comes to trace metals; these elements play crucial roles in biological processes \citep{DSW01,WDE04}. For example, molybdenum (Mo) is not only vital for biological functions such as nitrogen fixation \citep{Hil02,WDS02,SMR09} but has also been implicated in the origin of life itself \citep{SVP12,SVA13}. With regard to the latter, numerous experiments in recent times have demonstrated that metals (e.g., iron, nickel, chromium) in some form can serve as nonenzymatic catalysts for initiating protometabolic networks conceivably at the heart of life's origins \citep{MVM20,PAB20}.

To continue with our focus on molybdenum, it is possible that, by virtue of its high density either in elemental or compound form, it would be liable to sink downward and thence become depleted over time. We have commented on the prospects for an aerial biosphere on Venus in Section \ref{SSecModDef}, and the habitable region under consideration appears to have nutrients like phosphorus and sulfur at potentially adequate concentrations to support Earth-based microbes \citep{MTL21}, although significant uncertainties remain due to the paucity of reliable data \citep{CHJ21}. This optimism must be appropriately counterbalanced by the fact that the abundances of most trace metals are wholly unconstrained. Some of them (e.g., molybdenum) might not occur at desired abundances in the cloud decks of Venus, and perhaps other aerial habitable environments \citep{ML18}.

The second difficulty we wish to foreground has to do with the challenge of staying afloat in the habitable region, i.e., preserving the altitude. The net downward force (weight minus buoyancy) that would be experienced by a hypothetical organism of volume $\mathcal{V}_\mathrm{org}$ is given by
\begin{equation}
    F_d \approx \left(\rho_\mathrm{org} - \rho_a\right) \mathcal{V}_\mathrm{org} g,
\end{equation}
where $g$ is the acceleration due to gravity and the two densities were defined in Section \ref{SSecRateIssue}. If the organism is to avoid downward acceleration, which would eventually drive it beyond the habitable environment, the above force must be balanced by the drag delineated in (\ref{FDrag}). As long as the Reynolds number is not smaller than unity, we end up with the following scaling for the terminal velocity ($v_t$):
\begin{equation}\label{vt1}
    v_t \propto \left(\frac{\rho_\mathrm{org}}{\rho_a} - 1\right)^{1/2} \left(\frac{\mathcal{V}_\mathrm{org}}{\mathcal{A}_\mathrm{org}}\right)^{1/2},
\end{equation}
and if we specialize to a Reynolds number of order unity and lower, we arrive at
\begin{equation}\label{vt2}
    v_t \propto \left(\frac{\rho_\mathrm{org}}{\rho_a} - 1\right) \left(\frac{\mathcal{V}_\mathrm{org} \mathcal{L}_\mathrm{org}}{\mathcal{A}_\mathrm{org}}\right),
\end{equation}
in which $\mathcal{L}_\mathrm{org}$ represents the characteristic length scale of the organism. This equation is derived after invoking the Stokes relationship $C_D \approx 24/\mathrm{Re}$ \citep[Section 7.8]{TEF95} in (\ref{FDrag}), where $\mathrm{Re}$ is the Reynolds number; the parameter $\mathcal{L}_\mathrm{org}$ is manifested through $\mathrm{Re}$.

Our goal is to decrease the terminal velocity, as otherwise, the organism would exit the habitable region swiftly and thus lose functionality. Alternatively, this terminal velocity would need to be balanced by convection that is vigorous enough to counterbalance $v_t$. As revealed by (\ref{vt1}) and (\ref{vt2}), there are two noteworthy avenues whereby the terminal velocity can be lowered. First, this result is effectuated when $\rho_\mathrm{org} \sim \rho_a$, implying that the organism would possess a balloon-like structure primarily composed of air, analogous to the ``floaters'' envisioned by \citet{SS76}. In that event, however, the protection conferred against galactic cosmic rays, stellar energetic particles, and micrometeorites could be reduced owing to the relatively thin membrane enclosing air (and internal organs).

In addition, if the organism were to rely upon the aforementioned strategy, the majority of its area may be expected to consist of the balloon-like component. This surmise leads us naturally to the second approach wherein $\mathcal{A}_\mathrm{org}$ is substantially increased, while the volume, mass, and other properties of the organism are held fixed; boosting $\mathcal{A}_\mathrm{org}$ will suppress $v_t$ as seen from (\ref{vt1}) and (\ref{vt2}). Lifeforms in this vein could resemble pancakes or something similar, either with or without the balloon-like structure. However, as we shall describe below, this route of enhancing $\mathcal{A}_\mathrm{org}$ also has some prominent shortcomings associated with it.

An organism at temperature $T_\mathrm{org}$ will radiate away energy at the rate $P_\mathrm{org}$, which is expressed as
\begin{equation}
    P_\mathrm{org} = \sigma \mathcal{A}_\mathrm{org} T_\mathrm{org}^4,
\end{equation}
when we model it as a blackbody. Hence, if $\mathcal{A}_\mathrm{org}$ is significantly increased, it follows that $P_\mathrm{org}$ will be raised commensurately, thereupon leading to conceivably substantial energy losses. To counter this issue, it would be necessary for the organism to locate, acquire, and process raw materials for metabolism in greater quantities or at higher efficiency \citep{PV08}, both of which might engender additional barriers to the evolution of complex, motile, and macroscopic multicellular lifeforms.

Second, if the cross-sectional area is enlarged, the ratio of this quantity to the volume (denoted by SA:V) is raised by the same amount because the volume is held constant. While increased values of SA:V may be rendered advantageous in some respects, it is simultaneously accompanied by crucial downsides. For example, high SA:V ratios could translate to elevated rates of water loss \citep{KBS17}, which would be a critical issue in arid settings like the clouds of Venus \citep{HKD21}. Moreover, when the SA:V ratio becomes substantial, it might amplify: (a) the leakage rates of nutrient and useful metabolites, (b) the energy expended to sustain homeostasis, and (c) the costs of maintaining the structures (e.g., cells and tissues) that demarcate the organism \citep{BAB09,JGO13}; other negative ramifications of high SA:V can be gleaned from \citet{Koo00}.

\bibliographystyle{aasjournal}
\bibliography{LandOceansLife}

\end{document}